%% file: main.tex
\documentclass[conference]{IEEEtran}
\usepackage{cite}
\usepackage{graphicx}
\usepackage{url}
\usepackage{tikz}
\usepackage[utf8]{inputenc}
\usepackage[frozencache]{minted}
\AtBeginEnvironment{listing}{%
  }
\setminted{fontsize=\footnotesize,linenos, xleftmargin=1em, numbersep=0.5em}
\usetikzlibrary{positioning, calc, quotes}
\usepackage{moeptikz}
\graphicspath{ {images/} }

\begin{document}
\title{Rule-Based Translation of Application-Level QoS Constraints into SDN Configurations for the IoT}

\author{\IEEEauthorblockN{Jan Seeger\IEEEauthorrefmark{1}\IEEEauthorrefmark{2}, Arne Bröring\IEEEauthorrefmark{2}, Marc-Oliver Pahl\IEEEauthorrefmark{1}, Ermin Sakic\IEEEauthorrefmark{1}\IEEEauthorrefmark{2}}
\IEEEauthorblockA{\IEEEauthorrefmark{1}
Technical University Munich, Munich, Germany,
\{seeger, pahl\}@net.in.tum.de}
\IEEEauthorblockA{\IEEEauthorrefmark{2}Siemens AG, Munich, Germany,
\{arne.broering, ermin.sakic\}@siemens.com}
}
\maketitle

\hyphenation{SEM-IoT-ICS}
\begin{abstract}
In this paper, we propose an approach for the automated translation of application-level requirements regarding the logical workflow and its QoS into a configuration of the underlying network substrate.
Our goal is to facilitate the integration of QoS constraints in the development of industrial IoT applications to make them more reliable.
We follow an approach based on two semantic models:
The first model allows to design the workflow of an IoT application and to express application-level QoS requirements on its interactions. 
The second model captures the configuration of a network and can be used as input to a north-bound interface of an SDN controller.
Finally, we make use of rule-based semantic reasoning to automatically translate from the application requirements into SDN parameters.

\end{abstract}

\begin {IEEEkeywords}
IoT, Semantics, SDN, QoS
\end{IEEEkeywords}

\section{Introduction}
\label{sec:intro}
\input{sections/introduction.tex}

\section{Background \& Related Work}
\label{sec:relatedWork}

\input{sections/related_work.tex}

\section{Modelling IoT Compositions and SDN}
\label{sec:sdn-model}
\input{sections/model.tex}

\section{Application-level QoS Constraints}
\label{sec:app-level-constraints}
\input{sections/app_level_constraints.tex}

\section{Implementation \& Evaluation}
\label{sec:impl-eval}
\input{sections/implementation.tex}

\section{Conclusions \& Future Work}
\label{sec:conclusions}
\input{sections/conclusions.tex}

\section*{Acknowledgment}
This work has been supported through the project SEMIoTICS funded by the European Union's Horizon 2020 research and innovation programme under grant agreement No. 780315.

\bibliographystyle{IEEEtran}
\bibliography{IEEEabrv,main}

\end{document}

%% file: sections/introduction.tex
The Internet of Things (IoT) is rapidly growing. It consists of network-enabled devices with sensors and actuators that improve our comfort or make us safer. The IoT extends  ubiquitous communication to the physical world~\cite{gubbi_internet_2013}. The amount of information provided by the IoT and the diverse ways to interact with the physical world are challenging. 
One approach that has become widely used for the creation of new applications is the concept of \emph{service composition}\cite{sheng_web_2014}. Service composition means encapsulating functionalities that are provided by devices in dedicated services and combining them on a higher level. The ability to combine services is popular in IoT deployments, as tools such as ``If This Then That''\footnote{\url{http://ifttt.com}} or Node-RED\footnote{\url{http://nodered.org}} show. Furthermore, service composition facilitates easier development of applications built atop the compositional APIs. Such simplification can become a key enabler towards an IoT \emph{app economy} \cite{pahl:HomeSys2013}.

Looking at the industrial or building automation domains, composition of services is comparatively more difficult, both because of a lack of standards for interoperable communication, and an inability of current composition systems to state non-functional requirements that automation systems have. Such non-functional requirements include latency or bandwidth constraints. While progress is being made on standardizing communication interfaces and protocols (by groups such as the Fairhair Alliance\footnote{\url{http://fairhair-alliance.org}} or the Open Mobile Alliance\footnote{\url{http://openmobilealliance.org}}), the specification of QoS requirements is not supported by the current service compositions for automation systems.

Automation systems, particularly in the building domain, are converging to a shared infrastructure to reduce operating costs and to promote integration with information and communication technology. These shared infrastructures no longer provide the guarantees of an isolated automation network, such as the guaranteed delivery time and available bandwidth. SDN technology can help with this. Through the centralized management of network elements, advanced QoS requirements can be enforced in the network from a central point.

There has been little research on incorporating the tools that SDN provides and considering the requirements of service composition concepts to realize IoT applications. This paper presents an approach that bridges the application-layer and network-layer perspectives, by describing application requirements and automatically translating them into network/SDN configurations using semantically-enriched models. This semantic enrichment enables machine interpretable resource descriptions and the automated matching of existing devices and services to defined compositions. 

We follow an approach based on two semantic models: the first model for designing the application workflow builds up on our previous work on IoT Recipes~\cite{seeger_running_2018} and extends it by adding the ability to attach application-level QoS requirements into configurations of an SDN controller.
The second model developed here describes concrete SDN configurations. 
We further define a method for rule-based semantic reasoning that allows translating of high-level application-specific QoS constraints into lower-level SDN-specific QoS constraints, thus integrating QoS constraints with semantic application workflows.

%% file: sections/related_work.tex




    

A number of service composition platforms exist for composing automation tasks to new services.
A thorough survey on the field of cloud-focused QoS-aware web services composition can be found in \cite{hayyolalam_systematic_2018}.
The platforms described there are all cloud-focused, while we intend to model edge-level (i.e. local) service composition with QoS constraints.
For example, \cite{mokhtar_easy_2008} describes a QoS-aware service selection mechanism based on semantic matching.
Liu et. al. describe a reliable service composition platform in~\cite{liu_reliable_2017}, while Moustafa et. al. describe a stigmergic approach that qualifies provided QoS properties with trust \cite{moustafa_trustworthy_2016}. 
All of these approaches have in common that the QoS requirements are specified with the services themselves and the orchestration platform not taking the underlying network communication into account.
In contrast to the state of the art, by leveraging SDN functionalities, we aim to enforce QoS constraints on the network-level as well as the service level.

Software-defined networking (SDN) \cite{nunes_survey_2014} provides a fine-grain control of network settings.
It separates the control from the data plane and centralizes the control decisions on a single controller.
A typical mechanism to implement virtual topologies in SDN networks is using VLAN tags or OpenFlow flow-based traffic differentiation. Furthermore, OpenFlow can facilitate QoS-aware service differentiation by means of explicit queue assignment and per-flow metering mechanisms. 


Other established approaches for enabling QoS constraints on a per-application basis are Differential Services (DiffServ) \cite{blake1998architecture} and Integrated Services (IntServ) \cite{braden1994integrated}. DiffServ is a coarse-grained and decentralized approach for ensuring network traffic QoS.
However, its coarse class concept makes it unsuitable for expressing fine-grained QoS requirements for local automation systems.
IntServ architecture provides for fine-grained end-to-end support for QoS requirements.
It is however not widely supported by consumer hardware, and has scalability issues when it comes to larger systems.
Both of these protocols do not provide for a centralized view and control over the network, which complicates a global distribution of policies as we envision it.

Various research works focus on the enforcement of QoS parameters via SDN prootocols.
Naman et al.~\cite{naman_responsive_2018} describe the architecture for a network-exposed API that provides visibility into the network state, and an SDN assisted congestion control algorithm that utilizes network state information to achieve requirements that demand low latency and high bandwidth.
Akella et al.~\cite{akella_quality_2014} present a QoS-guaranteed approach for bandwidth allocation that satisfies QoS requirements for prioritized cloud users.
Kucminski et al.~\cite{kucminski_qos-based_2017} use a QoS-based routing scheme to prioritize important broadband data traffic over the less important one.
Li et al.~\cite{li_qos_2017} approach QoS guarantees by identifying the application at the SDN controller and setting up different QoS levels for different types of applications.
Guck et al.~\cite{guck2017detserv} develop a network model for guaranteeing latency bounds over standard network equipment with a reasonable runtime cost.
Gorlatch et al.\cite{gorlatch_enabling_2015} translate the high-level QoS requirement of response-time in real-time interactive applications to different types of network level latency requirements.

However, no systematic semantic modelling of QoS requirements has been attempted so far.
Our aim is the specification of a general model for the specification of QoS requirements on service compositions and as a translation target for application-specific QoS requirements.
This systematic model can then be used for providing functionality tools for enforcing service QoS requirements in an SDN enabled network.
t

%% file: sections/model.tex
\begin{figure*}[t]
   \centering
   \includegraphics[width=0.99\textwidth,trim={0 1.5cm 0 1.5cm},clip]{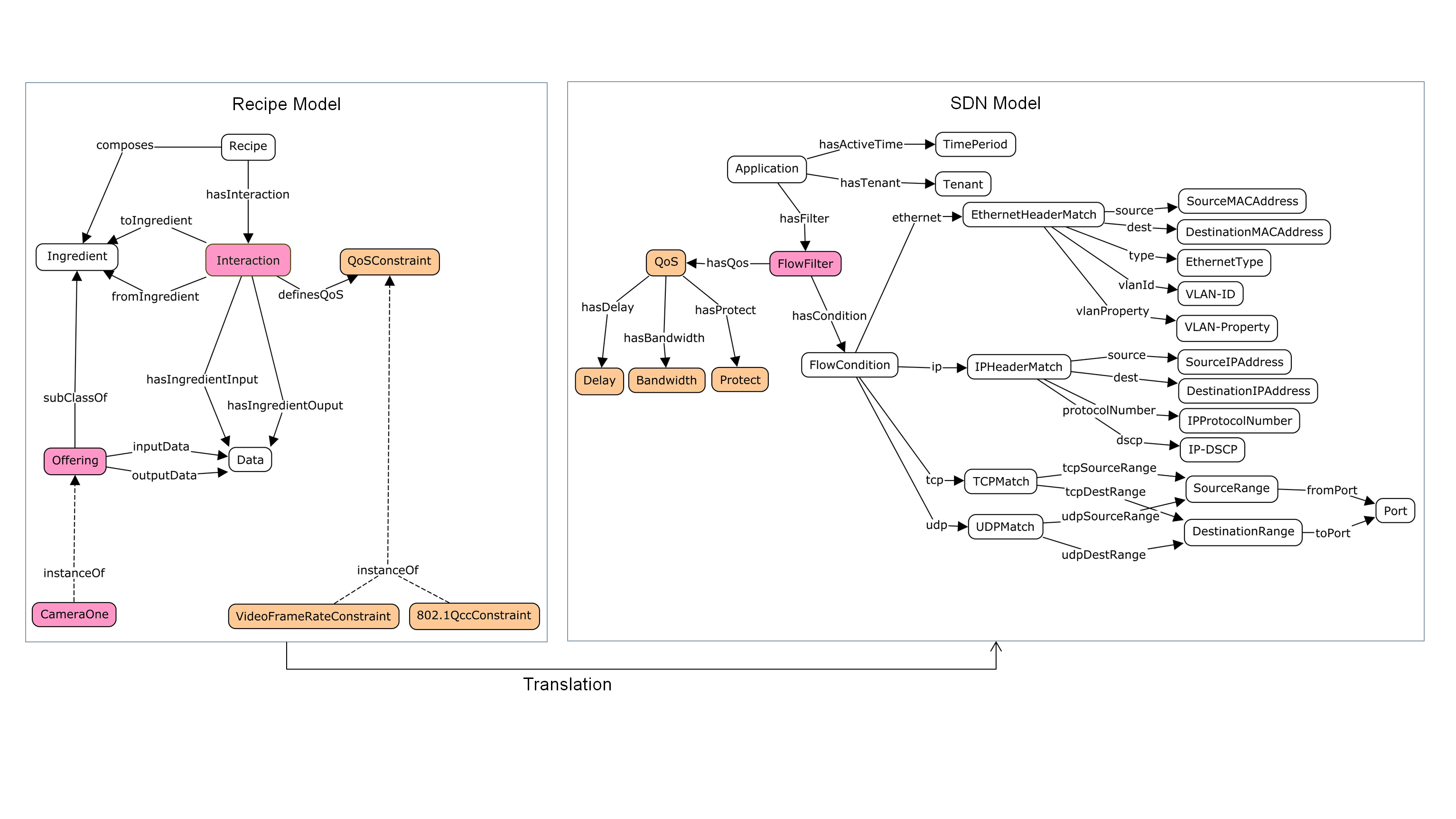}
    \caption{\label{fig:sdn-model}The two semantic models (as RDF triples) for IoT workflow compositions (left) and for defining SDN-based network configurations (right). Rules enable an automatic translation from recipe instances to SDN parameters.}
\end{figure*}



The semantic models of our approach are visualized in Fig.~\ref{fig:sdn-model}. They are defined as triples in the RDF format\footnote{https://www.w3.org/RDF/} and can be serialized e.g. in the N3 format\footnote{\url{https://www.w3.org/TeamSubmission/n3/}}.

On the left side of Fig.~\ref{fig:sdn-model}, the model to define abstract IoT compositions as \emph{recipes} is shown. This model is based on our previous work~\cite{seeger_running_2018, thuluva_recipes_2017}.
A recipe is a template for a workflow of interactions between multiple components, or \emph{ingredients}.
When a recipe is instantiated, ingredients are replaced with concrete components, which we call IoT \emph{offerings}.
An offering is a concrete service of an IoT device or platform that has inputs, outputs and a semantic category.
In this work, the recipe model is extended to allow the definition of application-level QoS constraints, which are then translated to SDN QoS constraints.
Therefore, the concept \textit{QoSConstraint} has been associated with an interaction of the recipe.
Based on this model, applications can be created in the form of a dataflow graph, as shown in the initial user interface design of Fig.~\ref{fig:sdn-interface}. Besides defining the interactions of the workflow, the user can specify constraints on the communication paths between devices.

SDN enables the enforcement and validation of QoS constraints on a service composition's network communication.
To take advantage of these tools, we need to model its parameters in a manner compatible with a service model.
We have chosen to model SDN concepts in a semantic fashion, for simplified integration with semantic service composition systems similar to the platform described in~\cite{seeger_running_2018}.

Our SDN model is depicted on the right side of Fig.~\ref{fig:sdn-model}. The design of this model is inspired by the data structures used by the northbound interfaces of SDN controllers, such as defined by \cite{sakic2018virtuwind}.
The central component of this model is the application. When the model is instantiated, this is the entry point to the definition of a specific SDN configuration. Associated with the application is a time period during which it is valid and a tenant who represents the user of the network. Every application is associated with an interface that comprises of the network node on which it runs as well as the physical port it is attached to.

A key concept associated with the application is the flow filter. 
Here, a destination (pointing to a specific interface), filter conditions, and QoS requirements are defined.
As QoS requirements, we have added delay, bandwidth and protect constraints.
This modelling is non-exhaustive, and depending on the functionality available at the store, more constraints can be added.
The delay constraint describes a maximum allowed latency between two endpoints, while the bandwidth constraint specifies a minimum guaranteed bandwidth between two endpoints. 
The protect constraint provides a mean to specify redundant packet transmission, which facilitates sending the same packet over different network links to improve the connection reliability.

These constraints are applied to flows that match the conditions attached to a single filter.
Currently, we included flow conditions to check for matches on the ethernet, IP, TCP and UDP protocols. Further protocols can be added, e.g., based on ARP addresses or ICMP packets.
As an example, to specify the maximum delay for a connection between a sensor and an actuator, we can instantiate a flow filter with a \emph{delay} QoS and a flow condition consisting of an IP header match with a source IP address of the sensor, and the destination IP address that of the actuator.
Then, the maximum delay constraint would be applied to all packets being sent from the sensor to the actuator.

Applications are the components that take advantage of the defined QoS constraint.
In the example user interface in Figure~\ref{fig:sdn-interface}, the shown recipe corresponds to the "tenant" concept. The tenant can have several independent applications.
Components in Figure~\ref{fig:sdn-interface} correspond to the "application" concept, where each application can have multiple QoS constraints.
Together, the applications (or components) realize a vision- and sound-based intruder alert function for an office building.


%% file: sections/app_level_constraints.tex
\emph{Application-level} QoS constraints refer to the possibility of defining such constraints on a high-level, independent of network-level specifics.
Application-level QoS constraints are thus an abstract description of an application's network requirements.
Due to being defined on the application level, such constraints are easier to define for the user, and can be stored independently of the specifics of the underlying network.
An example for the use and implementation of application-level constraints can be found in~\cite{gorlatch_enabling_2015}.

We have defined a scheme for expressing application-level QoS constraints as a collection of semantic rules.
Including these rules in the triple store together with the semantic models, the  application-level constraints are automatically translated by the semantic reasoner of the triple store into instances of the lower-level SDN model.
These instances can then be submitted as configurations to an SDN controller.

One use case for an application-specific constraint is constraining the \textit{frame rate} ($f$) for a camera stream that specifies the minimum frames/second the network needs to be able to transmit.
Since we define this constraint on the application level, information on the camera's data format and the resolution of the video stream is available to us.
If the video format's efficiency is $e \in [-\infty,1]$ and the video's resolution is $x \times y$, we can infer a minimum bandwidth with the calculation $\mathit{bw}=(1-e) * x * y * f$.
The bandwidth constraint derived from this equation can then be configured on the network.
If the application then changes (for example, switching to a video camera with a less effective video format), the application-level constraint can be reevaluated and changes can be applied to the network.

Another use case is the translation of 802.11Qcc\footnote{https://1.ieee802.org/tsn/802-1qcc/} traffic specifications into SDN requirements. The translation of a Qcc description with the maximum number of frames transmitted during a single interval as $N_F^{max}$ and the maximum length of transmitted frames as $W_F$ would be specified as $\mathit{bw} = N_F^{max} * W_F^{max}$.

\begin{listing}
  \centering
\begin{minted}{turtle}
:CameraOne a :Offering ;
    :resolutionX 1024 ;
    :resolutionY 786 ;
    :efficiency "0.8"^^xsd:float ;
    :address "192.168.178.25" .

:VideoFramerateConstraint
        a :Constraint ;
    :translatesInto [
        a :Calculation ;
        :targetConstraint :BandwidthConstraint ;
        :productOf (
            :resolutionX
            :resolutionY
            [
                a :Calculation ;
                :differenceOf (
                    1
                    :efficiency)
            ] , [
                a :ParameterValue ;
                :parameterRelation
                    :desiredFramerate ])] .

:VideoFramerateConstraintOne
        a :VideoFramerateConstraint ;
    :interactionFrom :CameraOne ;
    :interactionTo :ProcessingOne ;
    :desiredFramerate 20 .
\end{minted}
  \caption{\label{lst:device-constr-defin} Device and constraint definition in N3 format.}
\end{listing}

\begin{listing}
  \centering
\begin{minted}{turtle}
:productOf a :CalcFunction .
{
    ?calc a :Calculation ;
        :forConstraint ?constraint ;
        ?op ?list.
    ?op a :calcFunction .
    ?constraint :interactionFrom ?device .
    ?SCOPE e:findall (?value {
            ?rel list:in ?list .
            ?device ?rel ?value .
        } ?VALUES) .
    # Elided.
    (?VALUES ?CALCVALUES ?PARAMVALUES)
    list:append ?ALLVALUES .
    ?ALLVALUES e:length ?length .
    ?list e:length ?length .
} => {
    ?calc :inputValues ?ALLVALUES .
} .

{
    ?calc a :Calculation ;
        :productOf ?something ;
        :inputValues ?list .
    ?list math:product ?value .
} => {
    ?calc :hasResultValue ?value .
} .
{
    ?constraint a :Constraint ;
        :translatesInto [
            a :Calculation ;
            :hasResultValue ?value ;
            :targetConstraint ?sdnconstraint] ;
        :interactionFrom [
            a :Offering ;
            :address ?fromDeviceAddress] .
    # Elided.
} => {
    ?constraint :translatesTo [
        a ?sdnconstraint ;
        :hasValue ?value ;
        :matchFlow [
            a :FlowFilter ;
            :matchFromIP ?fromDeviceAddress ;
            :matchToIP ?toDeviceAddress]].
} .
\end{minted}
  \caption{\label{lst:appl-level-constr} Translation rules for a camera frame-rate constraint in N3 format.}
\end{listing}

Listing~\ref{lst:device-constr-defin} contains an example definition of a camera framerate application constraint and a device that this constraint can be applied to.
The definition of \texttt{CameraOne} contains the information necessary for calculating the constraint (resolution, efficiency, and address).
This information is stored in the orchestration system, and used at instantiation time of the recipe.
The definition of the constraint describes the translation from the high-level application constraint to lower-level network constraints.
In this case, our translation target is a \texttt{BandwidthConstraint}.
The target value of the bandwidth constraint should be calculated as per the use case defined above, where the bandwidth of the link is $x * y * (1-e)$  .
Application constraints can also translate into multiple network-level constraints.

Listing~\ref{lst:appl-level-constr} contains an excerpt of the translation implementation using the EYE reasoner\cite{verborgh_drawing_2015}.
The implementation takes the form of rules that are expressed as implications.
When the premise of the rule (the part before the $\Rightarrow$) holds, the conclusion of the rule is inserted into the triple store, with all existential variables (those prefixed with a '?') replaced with the bindings from the rule's premise.
Line 1 defines the \texttt{productOf} property as a calculation function that is resolved by the rule system.
The rule in lines 2--19 results in the recursive calculation of calculation values.
We do this by iterating over all the values in the argument list of the calculation relation (for example, \texttt{productOf}) and attaching the calculated values to the calculation.
The argument list can contain three types of values: Literals, which are used as-is, device properties, which are resolved from the device the constraint is applied to, and parameters, which are resolved from the constraint itself.
When all input values for a calculation are available (line 16), they are appended into a single list and attached to the calculation node.
Then, the calculation rule on lines 21 to 28 fires and computes the result using the reasoner's built-in \texttt{math:product} predicate.
This rule is replicated for other calculation instructions, such as \texttt{differenceOf} or \texttt{sumOf} (not included here).
When a result value for the root of the calculation has been computed, the rule in lines 30--47 generates the target constraint with the correct value.
Additionally, flow filter information from the device is used to generate a flow filter.
We can define QoS constraints for audio streams in a similar manner to realize the audio bitrate constraint in Fig.~\ref{fig:sdn-interface}.

The concept of application-level QoS descriptions harmonizes well with frameworks that support the abstract specification of compositions, such as COCOA~\cite{ben_mokhtar_cocoa_2007}, where abstract service compositions are treated as state machines, or the Recipe system from~\cite{seeger_running_2018}.

%% file: sections/implementation.tex
\begin{figure}[t]
   \centering
   \includegraphics[width=0.40\textwidth]{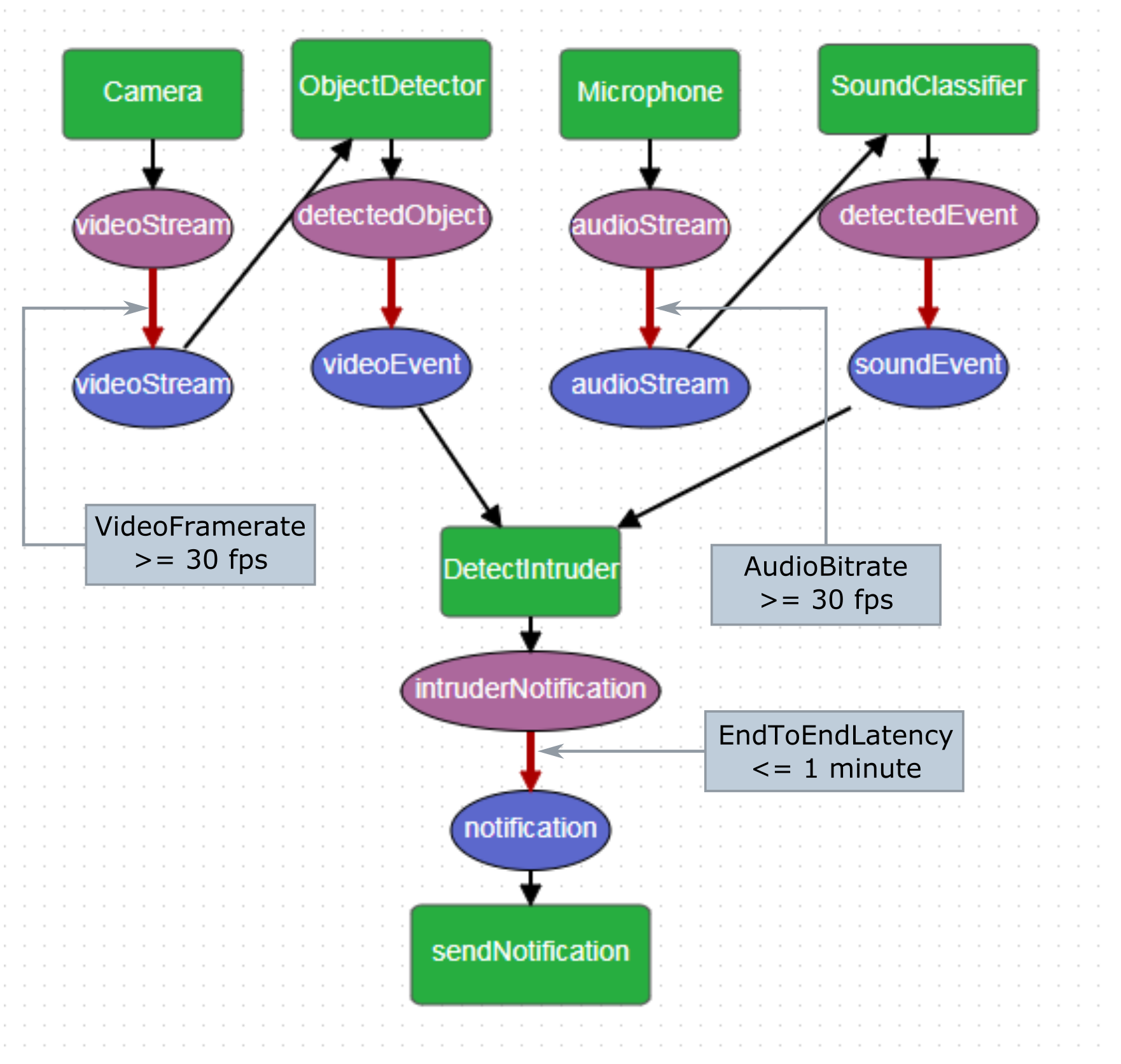}
    \caption{\label{fig:sdn-interface}User interface to configure network QoS on the application-level within a recipe defining an IoT service composition.}
\end{figure}

An example for a user interface (UI) design for the specification of application workflows and constraints can be seen in Fig.~\ref{fig:sdn-interface}.
This UI is based on our previous work in~\cite{thuluva_recipes_2017}.

In this example, the UI has been used to define a recipe that combines multiple services of devices in an intrusion detection system. For the camera and audio streams, the analysis services need a minimum amount of data to work correctly. To guarantee this, the user has specified application-level QoS requirements on the interactions between sensors and analysis services. Additionally, when a notification is generated by the system, it should be sent quickly. Otherwise, the intruder will be long gone when the notification is sent. For this, another constraint is attached that specifies a certain maximum time for a message to be delivered from one service to the next.

The abstract service composition and associated QoS constraints are first defined in the UI\@.
The user can then trigger the storing of the designed recipe as RDF triples (according to the above defined semantic model) in a triple store associated with the UI\@.
This results in semantic information in the triple store similar to that in Listing~\ref{lst:device-constr-defin}, however, without the \texttt{interactionFrom} and \texttt{interactionTo} parameters, since the recipe is still abstract.
When concretizing the recipe with specific components later, the \texttt{interactionFrom} and \texttt{interactionTo} properties are added to the constraint, which automatically starts the translation of the application constraint into concrete network constraints.
An external process regularly interrogates the triple store about all existing SDN-level constraints, and converts them to a format suitable for the targeted SDN controller, and send them to the SDN controller.
This enables the treatment of recipes containing application-level QoS requirements as QoS enabled applications that can be instantiated automatically using different concrete components.

\begin{figure}
\centering
\includegraphics[width=.45\textwidth]{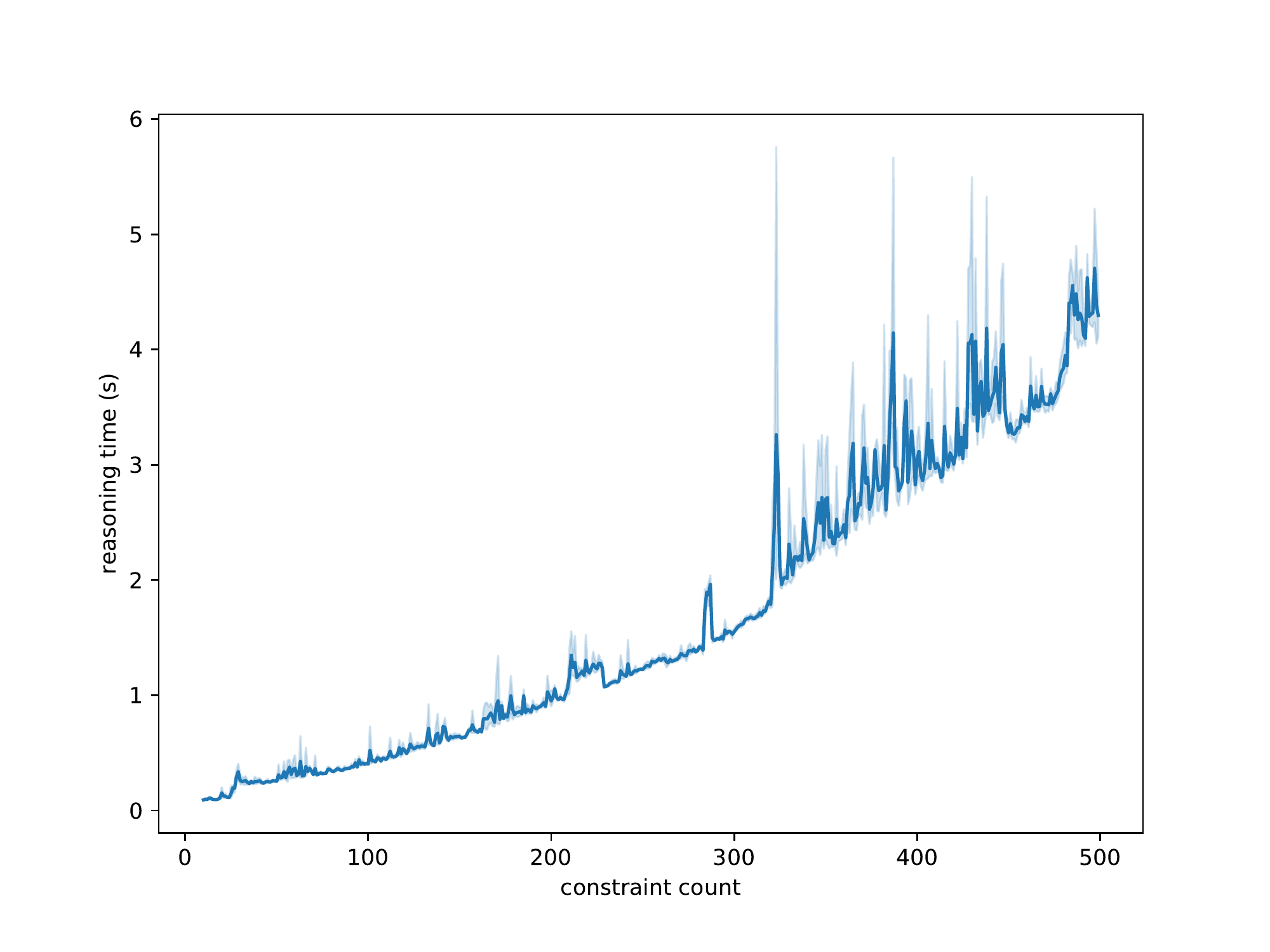}
\caption{\label{fig:performance} Number of constraints vs. required time to translate constraints with 100 devices. Each translation was run 5 times.}
\end{figure}

We have evaluated the performance of the translation of QoS constraints by repeatedly instantiating the "camera" constraint shown in Listing~\ref{lst:device-constr-defin} with 100 devices, and measuring the reasoning time. The results can be seen in Figure~\ref{fig:performance}. As expected, the Prolog-based reasoner performs efficiently with reasoning for 100 devices and 500 constraints taking less than 5 seconds on a 2.6 GHz 2-core virtual machine with 1 GB of RAM.

%% file: sections/conclusions.tex
In this paper, we have described a semantic model for defining SDN QoS constraints, and the use of this model in the instantiation of abstract service compositions.
Additionally, we have illustrated how application-level constraints (e.g., a video stream's frame rate, or a message's timeliness) can be translated into the provided model.
We have implemented this translation using a rule-based approach with the EYE reasoner.
This abstraction and the ability to define such constraints on the application-level supports application developers, as they do not have to know about networking details. I.e., we achieve flexibility and ease-of-use when defining service compositions with QoS requirements.

Modelling further application-level constraints will be done in future work, as it strongly depends on specific use cases.
In future, we will implement the abstract modelling approach in our recipe system~\cite{seeger_running_2018}. We plan to further elaborate the presented semantic model to a full ontology that enables the machine-interpretable definition of SDN configuration descriptions. 
Further, we will evaluate the ability of the system to run reliable service orchestrations.
This will involve the implementation of the user interface sketched in Fig.~\ref{fig:sdn-interface} and the implementation of an SDN management system to enforce those constraints in the network.